\documentclass{PoS}

\title{Thermodynamics of SU(3) gauge theory at fixed lattice spacing}

\ShortTitle{Thermodynamics of SU(3) gauge theory at fixed lattice spacing}

\author{\speaker{T. Umeda}$^a$, S. Ejiri$^b$, S. Aoki$^{a,c}$, 
        T. Hatsuda$^d$, K. Kanaya$^a$, Y. Maezawa$^e$, H. Ohno$^a$
        (WHOT-QCD Collaboration) \\
       $^a$ Graduate School of Pure and Applied Sciences, University of
       Tsukuba, Tsukuba, Ibaraki 305-8571, Japan \\
       $^b$ Physics Department, Brookhaven National
       Laboratory, Upton, New York 11973, USA\\
       $^c$ RIKEN BNL Research Center, Brookhaven National Laboratory, Upton, 
       New York 11973, USA\\
       $^d$ Department of Physics, The University of Tokyo,
       Tokyo 113-0033, Japan\\
       $^e$ En'yo Radiation Laboratory, Nishina Accelerator Research
        Center, RIKEN, Wako, Saitama 351-0198, Japan\\ 
        E-mail: \email{tumeda@het.ph.tsukuba.ac.jp}}

\abstract{We study thermodynamics of SU(3) gauge theory at
fixed scales on the lattice, where we vary temperature by
changing the temporal lattice size $N_t=(Ta_t)^{-1}$.
In the fixed scale approach, finite temperature simulations are
performed on common lattice spacings and spatial volumes. Consequently,
we can isolate thermal effects in observables from other
uncertainties, such as lattice artifact, renormalization factor, and
spatial volume effect. 
Furthermore, in the EOS calculations, the fixed scale approach is able
to reduce computational costs for zero temperature subtraction and
parameter search to find lines of constant physics, which are
demanding in full QCD simulations.
As a test of the approach, we study the thermodynamics of the SU(3)
gauge theory on isotropic and anisotropic lattices.
In addition to the equation of state, we calculate the critical
temperature and the static quark free energy at a fixed scale.}

\FullConference{The XXVI International Symposium on Lattice Field Theory\\
		 July 14-19 2008\\
		 Williamsburg, Virginia, USA}

\begin{document}

\section{Introduction}

Since finite temperature ($T>0$) lattice QCD is performed on lattice
with a temporal extent $N_t=1/aT$, qualitative calculations at high
$T$ may require lower simulation cost than that at $T=0$. 
However quantitative systematic studies at $T>0$ need huge simulation
cost often more than that at $T=0$. 
Because such study requires $T=0$ calculations at wide range of
lattice scale. 
It is a reason why recent large scale thermodynamics calculations are
often performed with the Staggered type quark formulations
\cite{lat08}, 
which needs lower computational cost than that with Wilson type
formulations \cite{AliKhan:2001ek}, the domain wall and overlap quarks
are all the more costly.
To make matters worse, some Wilson type quarks sometimes cause some
problems at coarse lattice,
e.g. nonperturbative clover coefficient $c_{SW}$ is reliably determined
at $a\sim 0.1$ fm or finer \cite{Aoki:2005et},
and the domain wall quarks encounter with strong residual quark mass
effects at coarse lattice \cite{Chen:2000zu}.
In spite of the difficulties,
results at $T>0$ with the Wilson type quark formulations are
desired, because the Staggered type quarks suffer from problems of the
flavor symmetry violation and the rooted Dirac operators.

Therefore we propose an alternative fixed scale approach to study
thermodynamics of QCD, where we vary $T$ by
varying the temporal lattice size $N_t=(Ta_t)^{-1}$ instead of the
conventional fixed $N_t$ approach.
In the fixed scale approach, $T=0$ results are common for
each $N_t$ ($T$) simulations.
It may be able to reduce total simulation cost drastically.
Furthermore, common parameters (except for $N_t$) enable us to
investigate pure thermal effects of observables without obstacles
coming from changing lattice spacing and spatial volume effects.

In this report, we test the approach in the SU(3) gauge theory
on isotropic and anisotropic lattices.
Our lattice action and some details of the EOS calculation are given
in Sect.\ref{sect2.1} and Sect.\ref{sect2.2}. 
Results of EOS 
are presented in Sects.\ref{sect2.3} and \ref{sect2.4}. 
The $T_c$ and the static quark free energy are discussed in 
Sect.\ref{sect3} and \ref{sect4}.
We conclude in the last section. 

\section{Equation of state}
\label{sect2}

\subsection{T-integration method}
\label{sect2.1}

In the fixed scale approach, 
to calculate the pressure non-perturbatively, we propose a new method,
``the T-integral method'' \cite{Umeda:2008bd} : 
\begin{eqnarray}
\frac{p}{T^4} = \int^{T}_{T_0} dT \, \frac{\epsilon - 3p}{T^5}
\label{eq:1}
\end{eqnarray}
based on another thermodynamic relation valid
at vanishing chemical potential:
\begin{eqnarray}
T \frac{\partial}{\partial T} \left( \frac{p}{T^4} \right) =
\frac{\epsilon-3p}{T^4}. 
\end{eqnarray}
The initial temperature $T_0$ is chosen such that $p(T_0) \approx 0$.
Calculation of $\epsilon -3p$ requires the beta functions just at the
simulation point, but no further Karsch coefficients are necessary. 
Since $T$ is restricted to have discrete values,
we need to make an interpolation of $(\epsilon -3p)/T^4$ with respect
to $T$. 

Since the coupling parameters are common to all temperatures,
our fixed scale approach with the $T$-integral method has several
advantages over the conventional approach;
(i) $T=0$ subtractions can be done by a common $T=0$ simulation, 
(ii) the condition to follow the LCP is obviously satisfied, and 
(iii) the lattice scale as well as beta functions are required only at
the simulation point.  
As a result of these,
the computational cost needed 
for $T=0$ simulations is reduced drastically.

When we adopt coupling parameters from $T=0$ spectrum studies,
the values of $N_t$ around and below the critical temperature $T_c$
are much larger than those used in conventional fixed $N_t$ studies.
For example, at $a \approx 0.07$ fm, $T \sim 175$ MeV is achieved by
$N_t \sim 16$. 
Therefore, for thermodynamic quantities around and below $T_c$, 
we can largely reduce the lattice artifacts over the conventional approach,
with much smaller total computational cost.
This is also a good news for phenomenological applications of the EOS, since
the temperature achieved in the relativistic heavy ion collision
at RHIC and LHC will be at most up to a few times the $T_c$
\cite{Hirano:2008hy}. 
We note here that, as $T$ increases, $N_t$ becomes small and hence the
lattice artifact increases.  
Therefore, our approach is not suitable for
studying how the EOS approaches the Stephan-Boltzmann value in the
high $T$ limit. 

\subsection{Lattice action}
\label{sect2.2}

We study the SU(3) gauge theory with the standard plaquette gauge
action on an anisotropic lattice with the spatial (temporal) lattice
size and scale $N_s$ ($N_t$) and $a_s$ ($a_t$), respectively.  
The lattice action is given by 
\begin{eqnarray}
S &=& 
\beta \xi_0
\sum_x\sum_{i=1}^3
\left[ 1-\frac{1}{3}\mbox{Re}\mbox{Tr} U_{i4}(x) \right] +
\frac{\beta}{\xi_0} \sum_x\sum_{i>j=1}^3
\left[ 1-\frac{1}{3}\mbox{Re}\mbox{Tr} U_{ij}(x) \right]\\
& \stackrel{\rm def.}{=} & 3N_s^3 N_t \, \beta \left(  \xi_0 P_t +
\xi_0^{-1}P_s \right) 
\end{eqnarray}
where $U_{\mu\nu}(x)$ is the plaquette in the $\mu\nu$ plane
and $\beta$ and $\xi_0$ are the bare lattice gauge coupling and bare
anisotropy parameters. 
The trace anomaly is calculated by
\begin{eqnarray}
\frac{\epsilon-3p}{T^4} &=&
\frac{N_t^3}{N_s^3 \xi^3} \, a_s 
\left(\frac{\partial\beta}{\partial a_s}\right)_{\!\xi}
\left\langle \left(\frac{\partial S}{\partial\beta}\right)_{\!\xi}
\right\rangle \label{eq:6}\\
&=&
\frac{3N_t^4}{\xi^3}
\left\langle
\left( a_s \frac{\partial \beta}{\partial a_s}
\right)_{\!\xi}
\left[
\left\{ \frac{1}{\xi_0} P_s + \xi_0 P_t \right\}
-\frac{\beta}{\xi_0}
\left(\frac{\partial \xi_0}{\partial \beta}\right)_{\!\xi}
\left\{ \frac{1}{\xi_0}P_s - \xi_0 P_t \right\}
\right]
\right\rangle \label{eq:7}
\end{eqnarray}
with $\xi=a_s/a_t$ the renormalized anisotropy.
$a_s (\partial\beta/\partial a_s)$ is the beta function.
$\partial \xi_0/\partial \beta$ vanishes on isotropic lattices.

\subsection{EOS on isotropic lattice}
\label{sect2.3}

Our simulation parameters are listed in Table~\ref{tab:para1}.
On isotropic lattices, we calculate EOS on three lattices to study the
volume and lattice spacing dependences.  
The ranges of $N_t$ correspond to $T=210$--700 MeV for the sets i1 and
i2, and $T=220$--730 MeV for i3, 
to be compared with $T_c\sim 290$ MeV.
The set a2 will be discussed later.
The $T=0$ subtraction is performed with $N_t=16$ for i1 and
i2, and with $N_t=22$ for i3. 
We generate up to a few millions configurations using the
pseudo-heat-bath algorithm. 
Statistical errors are estimated by the Jackknife analysis. 
appropriate bin sizes, which strongly depend on $T$.  
Typically, bin size of a few thousands configurations are necessary
near $T_c$, while a few hundreds are sufficient off the transition
region.  

\begin{table}[tb]
\begin{center}
\begin{tabular}{c|cccccccc}
\hline
set & $\beta$ & $\xi$ & $N_s$ & $N_t$  &$r_0/a_s$ & $a_s$[fm] &
$L$[fm] & $a(dg^{-2}/da)$ \\
\hline
i1 & 6.0 & 1 & 16 & 3-10 & 5.35($^{+2}_{-3}$) & 0.093 & 1.5 & -0.098172 \\
i2 & 6.0 & 1 & 24 & 3-10 &5.35($^{+2}_{-3}$) & 0.093 & 2.2 & -0.098172 \\
i3 & 6.2 & 1 & 22 & 4-13 & 7.37(3)      & 0.068 & 1.5 & -0.112127 \\
\hline
a2 & 6.1 & 4 & 20 & 8-34 & 5.140(32) & 0.097 & 1.9 & -0.10704 \\
\hline
\end{tabular}
\end{center}
\caption{Simulation parameters on isotropic and anisotropic lattices.
On isotropic lattices, we adopt $r_0/a$ of \cite{Edwards:1997xf}, and 
the beta function of \cite{Boyd:1996bx}.
Anisotropic $r_0/a_s$ is from \cite{Matsufuru:2001cp}, and the beta
function is calculated in \cite{Umeda:2008bd}. 
Lattice scale $a_s$ and lattice size $L=N_s a_s$ are calculated with
$r_0=0.5$ fm.} 
\label{tab:para1}
\end{table}

\begin{figure}[tb]
\begin{center}
\resizebox{70mm}{50mm}{
\includegraphics{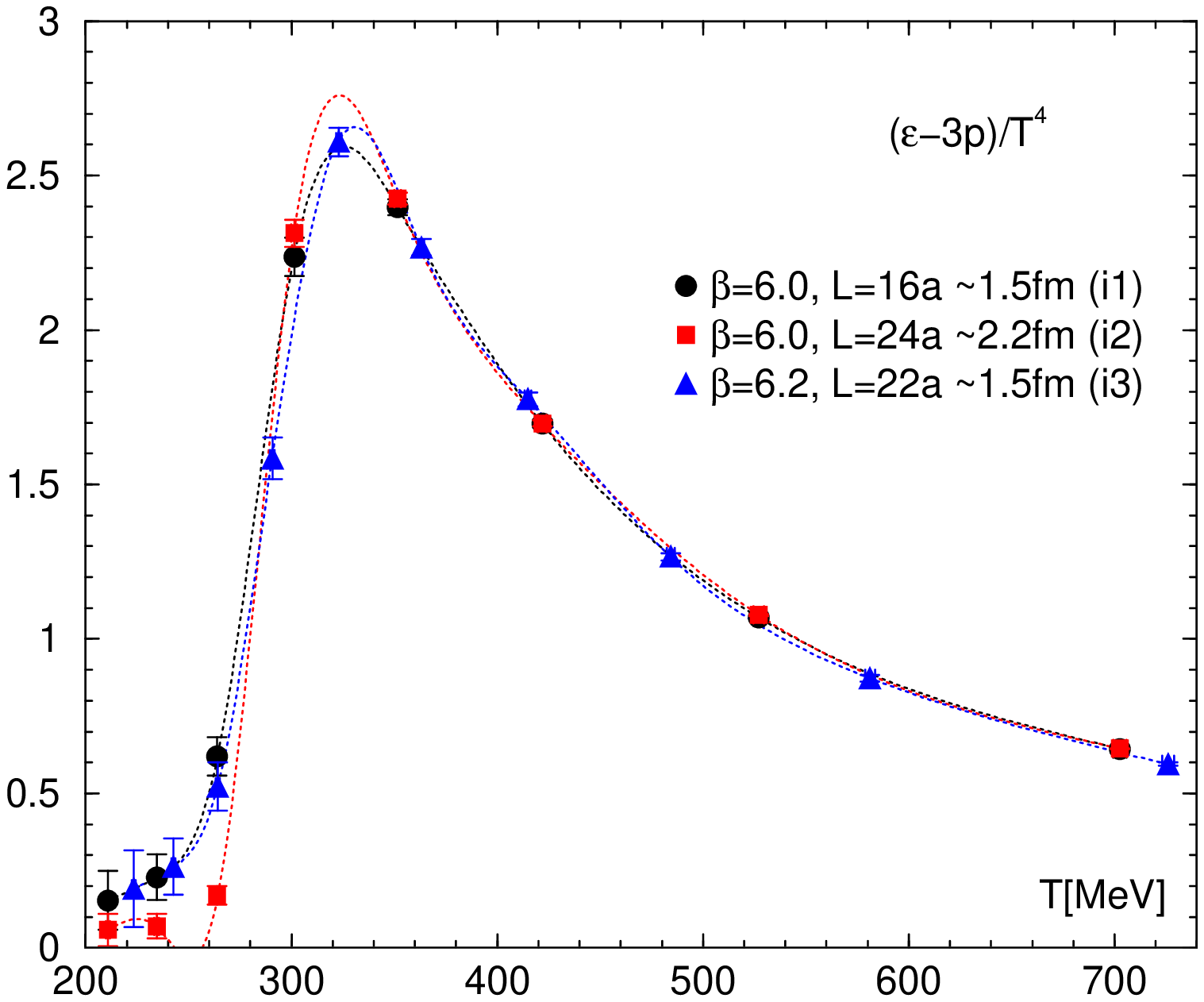}}
\hspace{2mm}
\resizebox{70mm}{50mm}{
\includegraphics{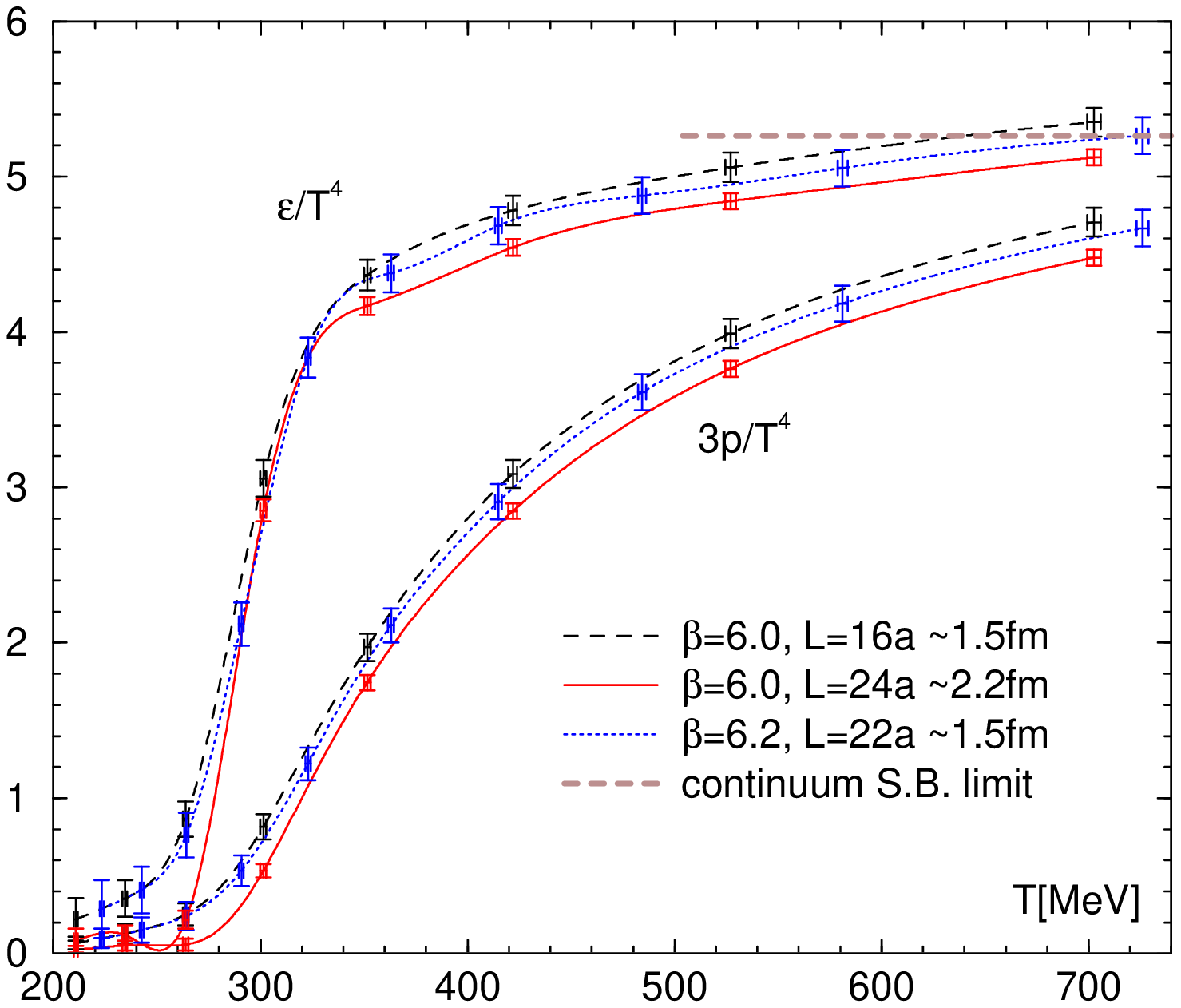}}
\end{center}
\caption{(Left) Trace anomaly on isotropic lattices. 
The dotted lines are natural cubic spline interpolations.
Horizontal errors due to the lattice scale are smaller than the symbols. 
(Right) The energy density and the pressure on isotropic lattices.}
\label{fig:iso1}
\end{figure}

Figure~\ref{fig:iso1} (Left) shows $(\epsilon-3p)/T^4$. 
Dotted lines in the figure are the natural cubic spline interpolations. 
At and below $T_c$, lattice size dependence is visible between the
sets i1 ($L\approx1.5$ fm) and i2 (2.2 fm). 
On the other hand, the lattice spacing dependence is small
between i1 ($a\approx0.093$ fm) and i3 (0.068 fm). 
At higher $T$, $(\epsilon-3p)/T^4$ on three lattices show good agreement. 
The integration of (\ref{eq:1}) is performed numerically using the 
natural spline interpolations shown in Fig.\ref{fig:iso1} (Left).
For the initial temperature $T_0$ of the integration, 
we linearly extrapolate the $(\epsilon-3p)/T^4$ data at a few lowest $T$'s
because the values of $(\epsilon-3p)/T^4$ at our lowest $T$ are not
exactly zero. 
In this study, we commonly take $T_0=150$ MeV as the initial
temperature which satisfies $(\epsilon - 3 p)/T^4 = 0$, and estimate
the integration from $T_0$ to the lowest $T$ by the area of the
triangle. 
Statistical errors for the results of integration are estimated by a
Jackknife analysis \cite{Okamoto:1999hi}.
Note that the error in the lattice scale do not affect the dimension
less quantity $p/T^4$. 

In Fig.\ref{fig:iso1} (Right), 
we summarize the results of EOS on isotropic lattices.
$\epsilon/T^4$ is calculated combining the results of $p/T^4$ and
$(\epsilon-3p)/T^4$.  
Since the lattice parameter dependence of $(\epsilon-3p)/T^4$ is
small except for the vicinity of $T_c$,   
we find that EOS has a similar shape except for the vicinity of $T_c$.
Near and below $T_c$, we observe a sizable finite volume effect
between $L \approx 1.5$ fm and $2.2$ fm, while the lattice spacing
effects are not so. 
At large $T$, we note a slight tendency that $p$ and $\epsilon$
decrease as the lattice size becomes larger and the lattice spacing
becomes smaller.  
Our results are qualitatively consistent with the previous results by
the conventional fixed $N_t$ method \cite{Boyd:1996bx}, but with much
reduced lattice artifacts around $T_c$ due to much larger $N_t$ there.  

\subsection{EOS on anisotropic lattice}
\label{sect2.4}

The anisotropic lattice with the temporal lattice finer than the
spatial one is expected to improve the resolution of $T$ 
without much increasing the computational cost.
To further test the systematic error due to the resolution of $T$, 
we perform the study with the $T$-integral method on an anisotropic lattice 
with the renormalized anisotropy $\xi = 4$. 
The simulation parameters are given as the set a2 in
Table~\ref{tab:para1}, which 
are the same as those adopted in \cite{Matsufuru:2001cp}.  
We vary $N_t=34$--8 corresponding to $T=240$--1010 MeV.
The $T=0$ subtraction is performed with $N_t=80$. 
We generate up to a few millions configurations. 
The beta function $a_s (\partial\beta/\partial a_s)_{\xi}$ 
with $\xi=4$ has calculated in our paper \cite{Umeda:2008bd},
and its value at our simulation point is given in Table~\ref{tab:para1}.
For $(\partial \xi_0/\partial \beta)_\xi$ 
we adopt the result of \cite{Klassen:1998ua}. 

\begin{figure}[tb]
\begin{center}
\resizebox{70mm}{52mm}{
 \includegraphics{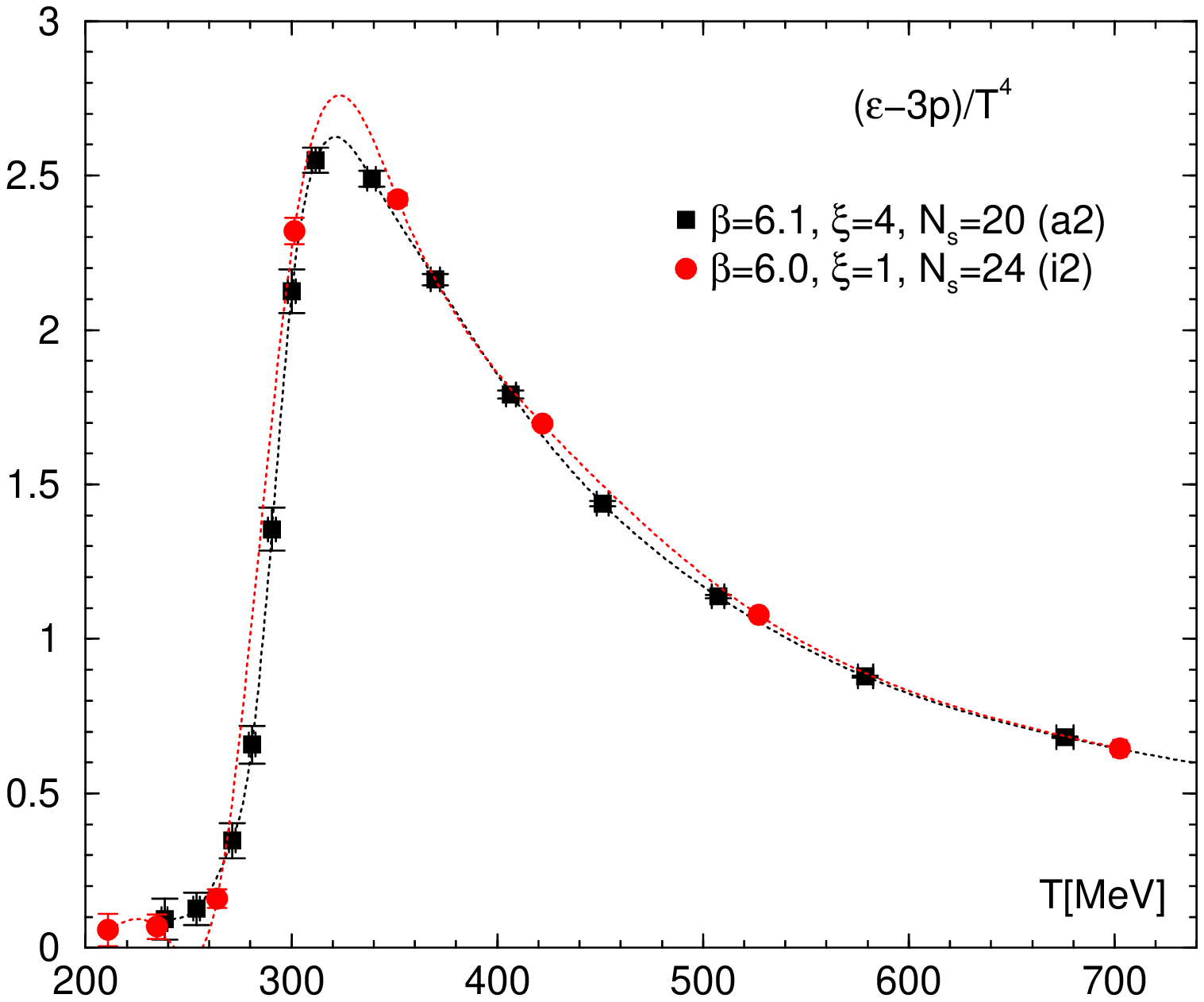}}
\hspace{2mm}
\resizebox{70mm}{50mm}{
 \includegraphics{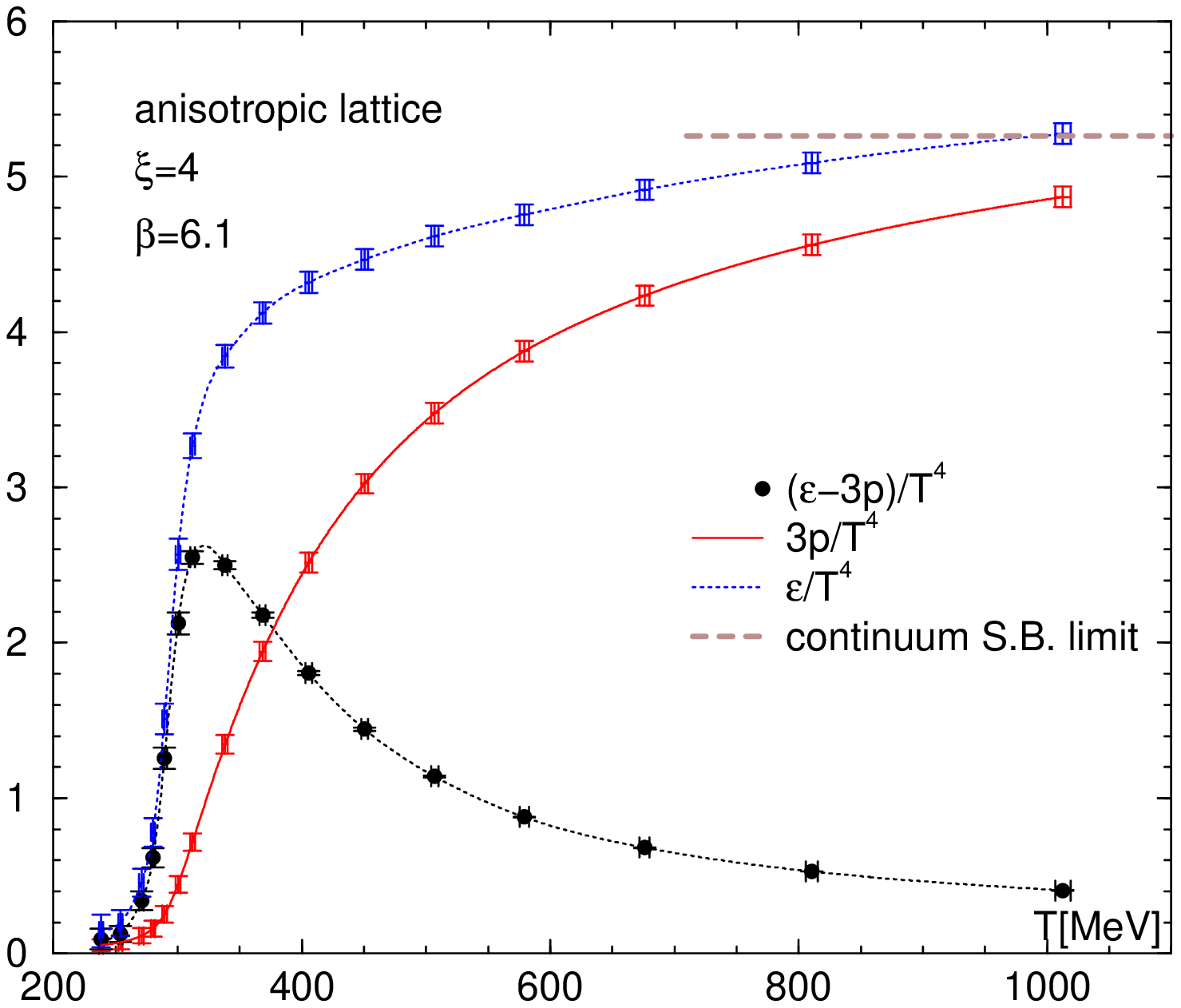}}
\end{center}
\caption{(Left) The trace anomaly on the anisotropic lattice (a2) and
the isotropic lattice (i2). Dotted lines show the cubic spline
interpolation. 
(Right) The EOS on the anisotropic lattice. } 
\label{fig:comp1}
\end{figure}

In Fig.~\ref{fig:comp1} (Left), we compare the trace anomaly obtained on the
anisotropic lattice with that on the isotropic lattice with similar
$a_s$ and $L$ (the set i2). 
We find that the results are generally consistent with each other
except for around $T_c$.
We note a systematic tendency that the trace anomaly on the
anisotropic lattice is slightly lower than that on the isotropic
lattice. 
According to this tendency, the pressure on the anisotropic lattice
is slightly smaller than that on the isotropic lattice at high $T$. 
%
The tendency may be understood by the smaller lattice artifact due to
the temporal lattice spacing on anisotropic lattices, since lattice
artifacts due to temporal lattice spacing are larger than that by the
spatial lattice spacings in thermodynamic quantities
\cite{Namekawa:2001}. 
%
Finally, we summarize our results of EOS on the anisotropic lattice in
Fig.~\ref{fig:comp1} (Right). 
We find that they are consistent with those on isotropic
lattices.

\section{Transition temperature}
\label{sect3}

Here we consider a possibility to compute the $T_c$
in the fixed scale approach. 
$T_c$ is determined by studying
temperature dependence of order parameters.
Strictly speaking, such temperature dependence should be
separated from other effects, such as renormalization, lattice artifacts, and spatial volume dependence. 
In the fixed scale approach, 
we can easily isolate the thermal effect on the observables.
On the other hand, the resolution of $T$ is restricted by descrete $N_t$.

\begin{figure}[tb]
\begin{center}
\resizebox{70mm}{50mm}{
 \includegraphics{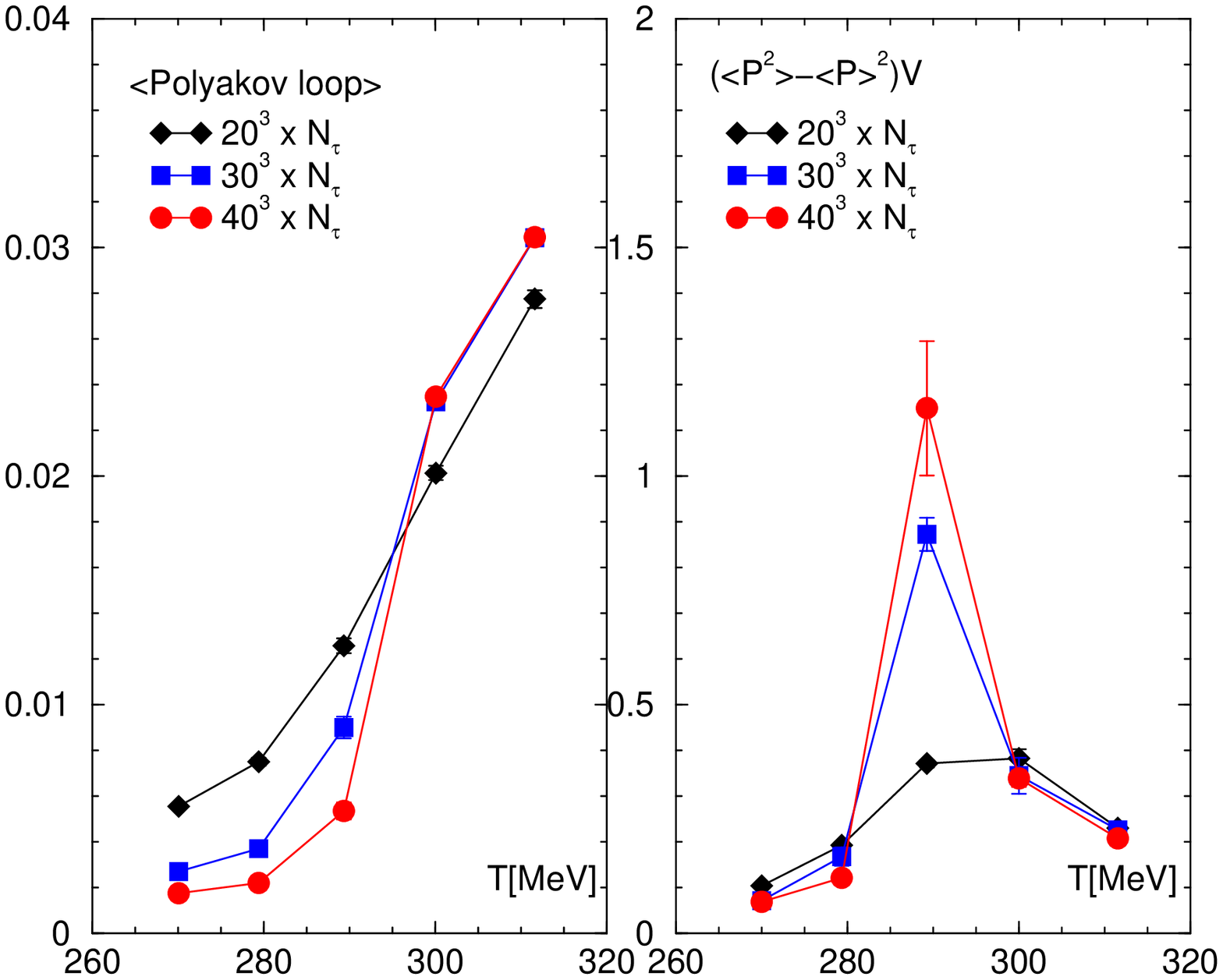}}
\hspace{2mm}
\resizebox{70mm}{50mm}{
 \includegraphics{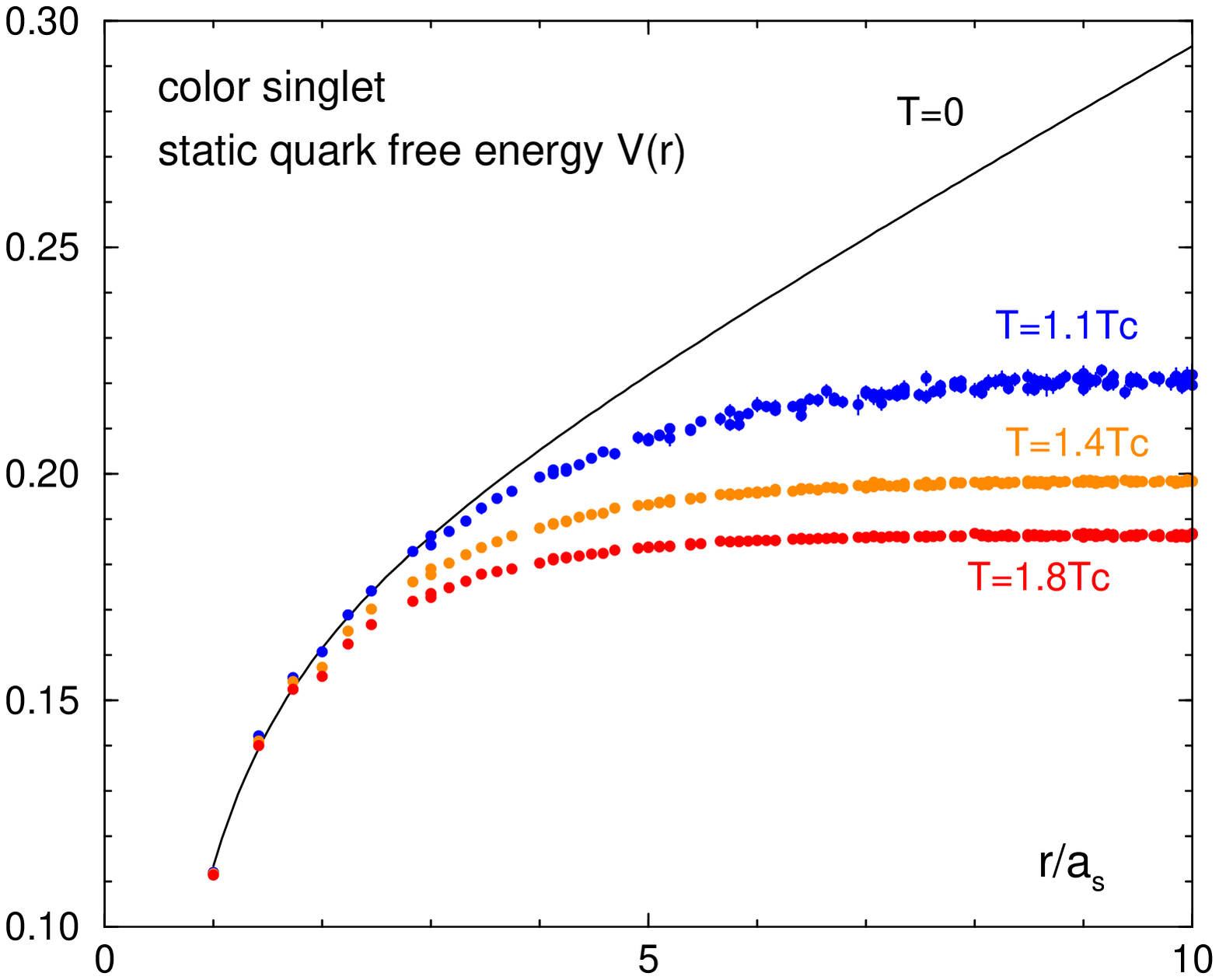}}
\end{center}
\caption{(Left,Center) Polyakov loop and its susceptibility on the
anisotropic lattice. (Right) Static quark free energy on the anisotropic lattice.}
\label{fig:tc}
\end{figure}

We calculate the Polyakov loop and its susceptibility on the
anisotropic lattice.
In addition to the lattices prepared for the EOS calculation, 
we generated different spatial volume lattices to study its
finite size scaling.
The parameters we adopted are listed in Table \ref{tab:para2}.
Since the SU(3) gauge theory has the global Z(3) symmetry, we calculate
the real part of Z(3)-rotated Polyakov loop and its susceptibility.
Our results are shown in the left and center pannels of Fig.\ref{fig:tc}.
Unlike the case of conventional fixed $N_t$ studies, the
renormalization factor is common to all temperatures.
From the susceptibility data we can find that the transition point locates
at $N_t \approx 28$ corresponds to $T_c=280$-300 MeV from the
$r_0$ scale setting. 
We also find that the peak height of the susceptibility increases with increasing the system volume,
in accordance with the 1st order nature of the transition.

\begin{table}[tb]
\begin{center}
\begin{tabular}{c|cccccccc}
\hline
set & $\beta$ & $\xi$ & $N_s$ & $N_t$  &$r_0/a_s$ & $a_s$[fm] &
$L$[fm] \\
\hline
a2-1 & 6.1 & 4 & 20 & 26-30 & 5.140(32) & 0.097 & 1.9  \\
a2-2 & 6.1 & 4 & 30 & 26-30 & 5.140(32) & 0.097 & 2.9  \\
a2-3 & 6.1 & 4 & 40 & 26-30 & 5.140(32) & 0.097 & 3.9  \\
\hline
\end{tabular}
\end{center}
\caption{Simulation parameters on anisotropic lattices
to study volume dependence of the susceptibility of the Polyakov loop.}
\label{tab:para2}
\end{table}

\section{Static quark free energy}
\label{sect4}

Finally, we study the static quark free energy $V(r)$ in the fixed scale
approach. 
In conventional fixed $N_t$ studies, the additive renormalization constant for $V(r)$ is diffrent  for each $T$ because the lattice spacing is different. 
Assuming that the short distance physics is independent of $T$, the constant term is conventionally adjusted by hand such that $V(r)$ around the smallest $r$ coinside with each other.

In the fixed scale approch, on the other hand, the common lattice spacing for all $T$ implies that the constant term in $V(r)$ should be common too.
Therefore, we can purely study the $T$ effects without adjusting the constant term.
In Fig.\ref{fig:tc} (Right) we show our results of $V(r)$ for the color singlet channel on the anisotropic lattice, without adjusting the constant term.
The solid curve in the figure is a fit result of $T=0$ potential. 
We find that $V(r)$ at different temperatures converge to a common curve at short distances.
We thus have confirmed 
the expectation 
that the short distance physics is independent of $T$.

\section{Conclusions}
\label{sect5}

We proposed a fixed scale approach to study the QCD thermodynamics on
the lattice.  
In this approach, $T$ is varied by changing the temporal lattice size
$N_t$ at a fixed lattice scale. 
To test the method, we applied it to the SU(3) gauge theory
on isotropic and anisotropic lattices. 
We found that the $T$-integral method to calculate the EOS works quite
well.
The main advantage of our approach is that the computational cost for
$T=0$ simulations, which are the most time consuming
calculations in the conventional fixed $N_t$ approaches, can be
drastically reduced.   
We may even borrow configurations of existing high precision
simulations at $T=0$. 
The approach is applicable to QCD with dynamical quarks too. 
We are currently investigating EOS in $2+1$ flavor QCD with
non-perturbatively improved Wilson quarks,  
using the configurations by the CP-PACS/JLQCD Collaboration
\cite{nf2p1-wilson}. 
With these fine lattices, the lattice artifacts around $T_c$
are much smaller than the conventional fixed $N_t$ approaches. 
We are further planning to use the PACS-CS configurations just at the
physical point \cite{PACS-CS}.


TU thanks H.~Matusufuru for helpful discussions and comments.
The simulations have been performed on supercomputers 
at RCNP, Osaka University and YITP, Kyoto University.  
This work is in part supported by Grants-in-Aid of the Japanese Ministry
of Education, Culture, Sports, Science and Technology
(Nos.~17340066, 18540253, 19549001, and 20340047). 
SE is supported by U.S.\ Department of Energy (DE-AC02-98CH10886).

\end{document}